# meV resolution in laser-assisted energy-filtered transmission electron microscopy

E. Pomarico[1], I. Madan[1], G. Berruto[1], G. M. Vanacore[1], K. Wang[2], I. Kaminer[2], F. J. García de Abajo[3,4], and F. Carbone[1*]

[1]*Institute of Physics, LUMES, École Polytechnique Fédérale de Lausanne (EPFL), Lausanne CH-1015, Switzerland*

[2]*Andrew and Erna Viterbi Department of Electrical Engineering, Technion–Israel Institute of Technology, 32000 Haifa, Israel*

[3]*ICFO-Institut de Ciencies Fotoniques, The Barcelona Institute of Science and Technology, 08860 Castelldefels (Barcelona), Spain*

[4]*ICREA-Institució Catalana de Recerca i Estudis Avançats, Passeig Lluís Companys, 23, 08010 Barcelona, Spain*

**The electronic, optical, and magnetic properties of quantum solids are determined by their low-energy (< 100 meV) many-body excitations. Dynamical characterization and manipulation of such excitations relies on tools that combine nm-spatial, fs-temporal, and meV-spectral resolution. Currently, phonons and collective plasmon resonances can be imaged in nanostructures with sub-nm and 10s meV space/energy resolution using state-of-the-art energy-filtered transmission electron microscopy (TEM), but only under static conditions, while fs-resolved measurements are common but lack spatial or energy resolution. Here, we demonstrate a new method of spectrally-resolved photon-induced near-field electron microscopy (SRPINEM) that allows us to obtain nm-fs-resolved maps of nanoparticle plasmons with an energy resolution determined by the laser linewidth (20 meV in this work), and not limited by electron beam and spectrometer energy spreading. This technique can be extended to any optically-accessible low-energy mode, thus pushing TEM to a previously inaccessible spectral domain with an unprecedented combination of space, energy, and temporal resolution.**

[*] Corresponding author: fabrizio.carbone@epfl.ch

In solids, low-energy many-body states emerge from the complex interplay between topology, electronic correlations, and structural dynamics.[1] Electron and photon-based inelastic scattering techniques can access this realm to provide information on energy scales and dispersion relations via spectroscopy[2], and on symmetry via selection rules.[3] When combined with microscopy, such as in energy-filtered electron microscopes, sub-nm spatial resolution can also be attained.[4] However, the energy resolution of these methods is fundamentally limited by the energy spread of the probe beam, which forbids access to features in the far-infrared spectral range. This limitation is further incremented by the finite energy resolution of the spectrometer.[5] For these reasons, much effort is currently underway to develop tools with improved energy resolution, simultaneously combined with space and time resolution, capable of resolving the structural, electronic and magnetic textures that characterize quantum solids and nanostructures.

A way to circumvent the energy-resolution problem is offered by laser-based pump-probe methods, in which low energy modes correspond to long-period oscillations that can be coherently excited by light pulses and directly probed in real time via spectroscopy[6,7] or microscopy.[8,9] Time-resolved methods further enable the determination of the lifetimes of specific excitations[10] and offer a unique handle to control the out-of-equilibrium properties of a solid.[11] Unfortunately, all-optical pump-probe techniques are limited by diffraction in far-field microscopes and by interaction with a sampling tip in near-field setups, yielding a spatial resolution of tens of nm at best.[12,13] Recently, pulsed electron beams have been combined with ultrafast lasers to simultaneously achieve nm-fs space-time resolution in the technique known as photon-induced near-field electron microscopy (PINEM),[14,15,16,17] successfully applied in mapping plasmon dynamics at surfaces,[9] interfaces,[18] and buried layers.[19] However, energy resolution in PINEM is obtained through an electron spectrometer, so the results are also affected by energy spread in the electron beam, which in practice limits the resolution to the eV range. Better energy resolution ~10 meV has been recently reported in state-of-the-art TEMs through electron energy-loss spectroscopy (EELS),[20,21,22] but with much poorer or no time resolution. The sought-after nm-fs-meV combined resolution is therefore remaining as an insurmountable challenge with currently existing approaches.

To go beyond these limitations, we have developed a disruptive approach that adopts tunable-wavelength light pulses for photo-excitation combined with energy-filtered TEM (Fig. 1, see details in Methods). A scheme of our experimental apparatus is depicted in Fig. 1a. We demonstrate the potential of our technique by applying it to study plasmon resonances (PRs) in long metal nanowires (NWs), driven by the strong potential of these excitations in a broad range of areas including optical sensing,[23] and more recently, optoelectronics at infrared energies in 2D materials.[24,13] The study of plasmons has a long tradition in electron microscopy[20,25] and should benefit from combined nm-fs-meV resolution.

In general, the transient electric field associated with the passage of swift electrons in the proximity of a nanostructure excites PRs, which are visible as features in an energy-loss spectrum (Fig. 1b). In time-resolved electron microscopy, the electron beam current is severely attenuated,[26] while the shortening of electron pulse duration produces broadening of the electron energy, which in practice stays at the eV level, thus resulting in significantly reduced contrast of the plasmon features (Fig. 1c). In contrast, in PINEM, a specific plasmon resonance can be strongly excited by illuminating the nanostructure with laser pulses at a resonant energy, resulting in the exchange of multiples of the PR energy with the imaging electron beam (Fig. 1d):

the external laser supplies photons in a coherent state, which consequently populate the plasmon also in a coherent state with average occupancy $n_o$>>1 for high light intensities; multiple plasmon gains and losses by the electron then take place with a probability proportional to $n_o$ times the spectral probability in conventional EELS.[27] For this reason, in PINEM, filtering the electron beam to image all inelastically scattered electrons yields the spatial profile of one specific PR, unlike conventional spectral imaging, which requires a filtering pass-energy tuned to the individual resonance. Ultimately, by scanning the laser wavelength across PRs and mapping their spatial profile in real space via PINEM imaging, one can resolve the modes with an energy resolution given by the laser linewidth (Fig. 1e), and thus no longer limited by the electron beam energy bandwidth.

In our spectrally-resolved PINEM (SRPINEM) method, we use a high-current electron beam ($\approx 10^2$ electrons per pulse) in combination with the photo-exciting light pulses. The energy distribution of such a beam is depicted in Fig. 2a before and after the arrival of the photo-excitation with 100 fs, 1.08 eV, 5 mJ/cm$^2$ pulses. The electron beam energy spread is as large as 6 eV due to intra-pulse electron-electron Coulomb repulsion.[26] Importantly, the overall spectrum changes when electrons and light pulses arrive in coincidence at the NW (Fig. 2a, red trace). A depletion is observed close to the zero of energy (referenced to the TEM acceleration voltage, 200 kV) corresponding to the zero-loss peak (ZLP) and accompanied by an increase of counts towards the tails of the spectrum. This is a typical fingerprint of the PINEM effect, which originates from inelastic scattering of the electrons by the photo-excited SPPs. Note that, as schematically depicted in Fig. 1c, the PINEM sidebands are not resolved because of the large energy bandwidth of the electron beam. However, by subtracting an image recorded before illumination (Fig. 2b) from one recorded in coincidence with the photo-excitation (Fig. 2c), the spatial distribution of the photo-induced plasmonic field is clearly retrieved (Fig. 2d). In this specific instance, Fig. 2d highlights the resonance mode of order $n$ = 16 (number of near-field nodes along the wire length) in a silver NW (100 nm diameter, 8 μm length) deposited on a $Si_3N_4$ membrane (see Methods).

We resolve the spectral lineshape of the PR supported by the silver NW by applying the described method to record a series of images as a function of the photo-exciting laser wavelength. When the photon energy is tuned to a PR (Fig. 3), a clear spatial profile is obtained, revealing the order of the mode $n$ through the number of PINEM intensity nodes along the wire (Fig. 3b). This is also evidenced by the spatial Fourier transform (FT) of the image, which allows us to retrieve the periodicity of the field and the strength of the resonance. By plotting the area of the FT peak associated with a particular PR as a function of the driving laser wavelength, we finally obtain the plasmon spectral profile with an energy resolution of 20 meV, only limited by the laser linewidth (Fig. 3c). Such a resolution is the result of the absolute Fourier limit of the laser pulse that cannot be bypassed, creating a spectral/temporal trade-off. In our current study, the laser system was prepared to deliver tunable near-infrared (near-IR) ~100 fs pulses with ~20 meV bandwidth, adapted to resolve the PR spectral shape and its temporal dynamics. To reach a finer energy resolution (less than 5 meV), 1-2 ps pulses have to be used, sacrificing time resolution. This could be of interest for example to study long-lived, sharp plasmons, such as those in high-quality graphene.[13]

Notably, this method allows us to combine nm-fs-meV energy-space-time resolution in a single experiment, yielding a complete characterization of a low-energy collective mode. In Fig. 4a, we show our measured spectra of the wire PRs in the range between 800 and 1080 meV (see supplementary figures for spatial profiles of all modes), highlighting the presence of *n* = 13-16 resonances. Incidentally, the optical excitation of both even and odd modes in the geometry of our experiment is due to the skew angle of the long wire relative to the electron beam direction.[9] The measured spectra are in good agreement with SRPINEM simulations based upon boundary-element-method calculations of the optical field (see Methods). We attribute the discrepancies in mode energies to imperfections of the NW, which result in a deviation from the ideal Fabry-Perot reflections calculated in a featureless wire, for which a more uniform mode spacing is predicted. In Fig. 4b, we show the transversal spatial profile with a lateral resolution of 8 nm along the direction perpendicular to the nanowire axis; the plasmonic field is observed to decay exponentially with a characteristic length of 280 ± 20 nm for *n*=13, roughly diminishing with the increasing order of the resonance, which is consistent with previous observations by spectral imaging[20] and also with an approximate transversal decay given by the plasmon wavelength divided by $2\pi$.[13] The temporal evolution of the plasmonic field is depicted in Fig. 4c, where we plot the intensity of the PINEM signal obtained with short electron bunches as a function of the time delay between laser and electron pulses. By assuming an instrument response function (IRF) of 250 ± 50 fs (see Fig. S2 of Ref. [19]), we obtain a plasmon lifetime of 90 ± 40 fs via a deconvolution procedure applied to the temporal trace.

We stress that this methodology is not limited by the energy broadening of the incident electron beam, in contrast to conventional inelastic scattering methods. In fact, the central frequency and the resolution necessary to image a specific mode are solely determined by the light excitation properties and not by the electron beam or spectrometer characteristics. This is an important advantage because light sources of high monochromaticity and controlled temporal profile are readily available at energies covering a wide spectral range, thus giving access to a broad parameter space compared with state-of-the-art electron optics. While in our current report we employ a fs infrared beam obtained via nonlinear wavelength conversion in an optical parameter amplifier, ultra-high resolution spectroscopy experiments (even below 1 meV if one partially sacrifices time-resolution) can be conceived using the intrinsic small linewidth of laser sources, or using spectral shaping techniques. Furthermore, our method can be applied to the investigation of arbitrarily low-energy modes, provided that they can be selectively excited by tunable light pulses. A clear candidate is phonons, which have been recently resolved in energy-filtered TEM[22] and can be photo-excited to manipulate the properties of quantum solids and nanostructures. By resolving the spectral, spatial, and temporal profiles of the light-induced phonon field in a nanostructure, one could simultaneously retrieve the dynamical coupling between the lattice and the electronic structure through lineshape analysis, real-space observation of the coherent structural modes, and anharmonic decay in real-time, yielding a unique tool to visualize and control the physics of advanced nanostructured materials.

## METHODS

**Sample preparation.** Silver nanowires with diameter between 50 and 100 nm and lengths of the order of 10 $\mu$m (Sigma-Aldrich, 739421) were dispersed in ethanol through sonication for 30 min. Samples were then prepared by dropcasting a single drop of the resulting suspension on 50 nm thick $Si_3N_4$ support films (Ted Pella, 21509-10), and were air dried for a couple of hours before examination in the ultrafast transmission electron microscope (UTEM, operated at 295 K and around $10^{-5}$ Pa).

**Experimental apparatus.** A 300 kHz train of linearly polarized, 800 nm, 80 fs light pulses with 10 $\mu$J per pulse was split to generate two beams. One of the beams was frequency tripled to deliver few-nJ ultraviolet pulses utilized to photoemit electrons from a truncated-cone $LaB_6$ cathode (15 µm diameter truncation plane, AP-Tech) in a modified JEOL JEM 2100 microscope operated at 200keV,[26] as schematically shown in Fig. 1a. The second 800 nm beam was used to seed an optical parametric amplifier (TOPAS HR, Light Conversion), producing signal beams between 1150 (1080 meV) and 1550 nm (800 meV), with energy per pulse in the 1-2 µJ range, depending on the wavelength. After passing through an optical delay line, the near-IR pulses were focused inside the UTEM to a spot size of around 40 µm, such that uniform photo-excitation in the field-of-view of the photoelectron beam was obtained. The electron beam was perpendicular to the sample (i.e. to the thin film substrate), although the NWs could still be tilted with respect to it.

For the PINEM experiments, a constant optical fluence of 5 mJ cm$^{-2}$ at all wavelengths was used. Polarization of the pump pulses was set by using a broadband half-wave plate in the near-IR spectral regime. Energy-filtering of the electrons was performed by a post-column Gatan Quantum GIF electron energy-loss spectrometer (GIF). A 2048 x 2048 pixel CCD camera operated with a dispersion setting of 0.05 eV per channel was used to detect the filtered electrons. Imaging was performed by using an electron beam with $\approx 10^2$ electrons per pulse and selecting a 4 eV window at the center of the ≈6 eV zero-loss peak obtained under these electron intensity conditions. Images of the wire in the presence of the pump pulse and before its arrival were obtained with 20 minutes integration and 5 minutes alternation to average over fluctuations of the electron beam intensity and sample position.

**Image processing.** Images of the investigated NW at all wavelengths with and without the photo-excitation were spatially aligned using a 2D cross-correlation algorithm, after correcting for the different overall amount of counts caused by temporal fluctuations of electron beam intensity and instrumental conditions. Spikes above a fixed threshold value were removed and single-pixel noise was median-filtered. Background was subtracted from every image after fitting its spatial distribution. Then, images of the photo-induced plasmonic fields were retrieved by taking the difference between the images recorded before and in coincidence with the photo-excitation. Spatial image calibration was performed by comparing the size of the nanowire with independent references.

**Numerical simulations.** The interaction between swift electrons (velocity $v$, moving along $z$) and nearly-monochromatic light (frequency $\omega$, electric field $E_z$ along $z$) in PINEM is quantified by the coupling integral $\beta = (e\gamma/\hbar\omega) \int dz\, E_z(z)\, e^{-i\omega z/v}$, where $\gamma = 1/\sqrt{1 - v^2/c^2}$.[28] The probability that the electron has gained or lost an energy $\ell\hbar\omega$ is $J_\ell^2(2|\beta|)$, so in particular, the depletion of the zero-loss peak ($\ell = 0$) is $1 - J_0^2(2|\beta|) \approx |\beta|^2$ for small $\beta$. We take this value as an estimate of the SRPINEM signal, calculated for each electron impact parameter relative to the wire, with the light electric field obtained via the boundary-element method[29] under external light illumination of a silver NW (100 nm diameter, 8 μm length, dielectric function from Ref. [30]). The wire is taken to be oriented at 45° from the electron beam direction.

**ACKNOWLEDGEMENTS**

The LUMES laboratory acknowledges support from the NCCR MUST of the Swiss National Science Foundation. E. P. acknowledges financial support from the Swiss National Science Foundation through an Advanced Postdoc Mobility Grant (P300P2_158473). The authors would like to acknowledge A. Cavalleri for useful discussions.

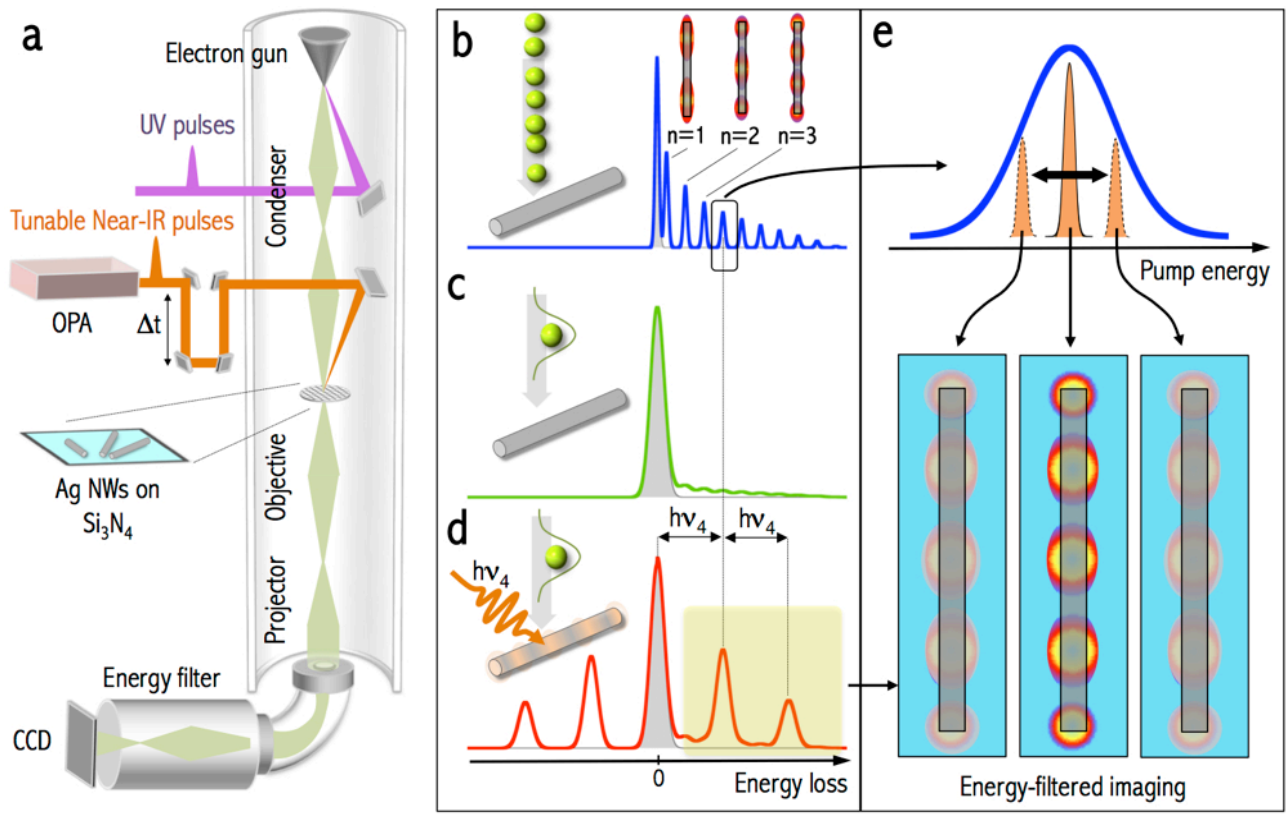

Fig. 1 **Spectrally-resolved photon-induced near-field electron microscopy: experimental setup and concept.** *a* Scheme of the experimental apparatus. *b* An EELS spectrum reveals multiple plasmon resonances in an individual nanowire, excited by the passing electrons and mapped with nm resolution by raster scanning the electron beam (plasmon map insets). *c* EELS spectrum of the nanowire under typical ultrafast energy-filtered microscopy conditions. The visibility of the electron-induced plasmon resonances is severely reduced. *d* Sketch of an EELS spectrum of the nanowire upon photo-excitation by light pulses of energy $h\nu_4$, tuned to the n=4 plasmon-resonance frequency. This specific mode exchanges several times its characteristic energy with the electrons. *e* Concept of our experiment: the laser excitation wavelength is scanned and the plasma resonance profile retrieved via quantitative analysis of the energy-filtered images.

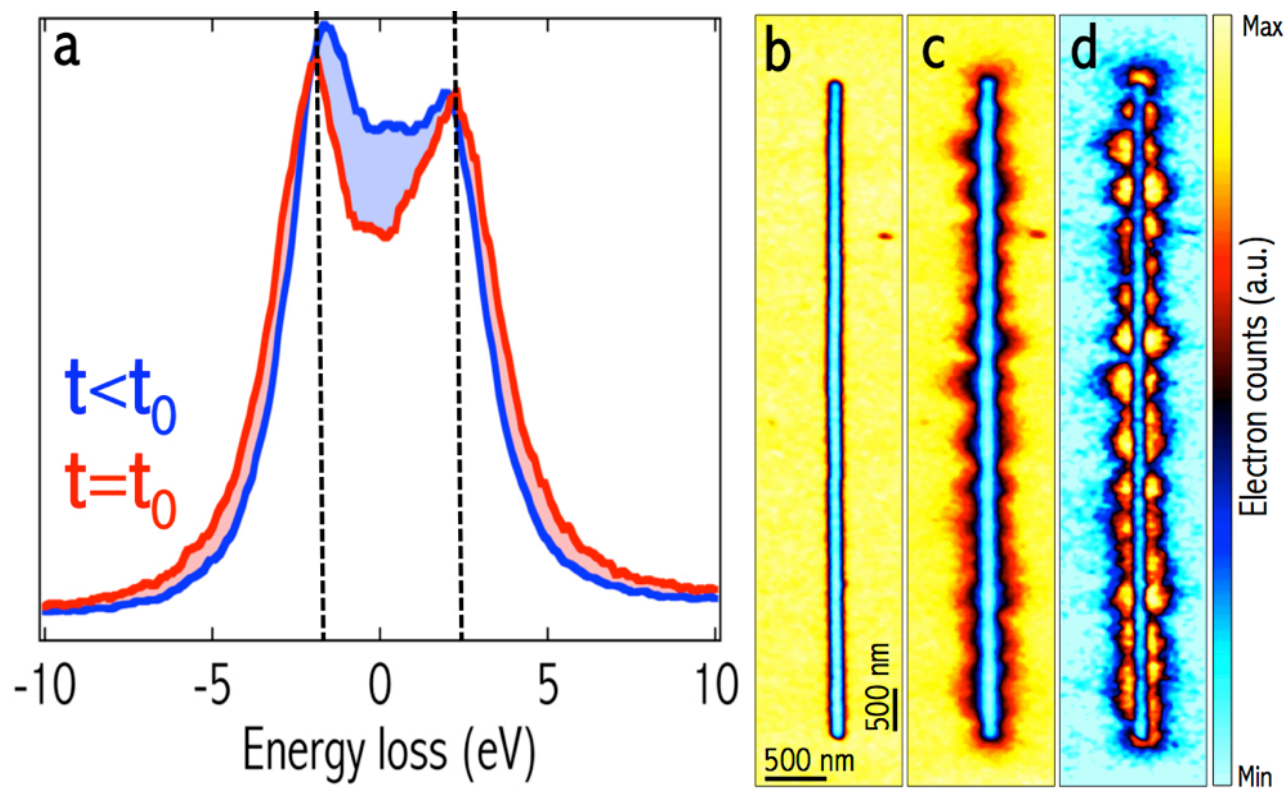

Fig. 2 **Plasmon imaging method. a** EELS spectrum of the nanowire taken with ≈$10^2$ electrons per pulse before (blue) and during (red) the arrival of light excitation at time $t_0$. **b,c** Nanowire images with (c) and without (b) light excitation. **d** Plasmon image obtained as the difference between b and c.

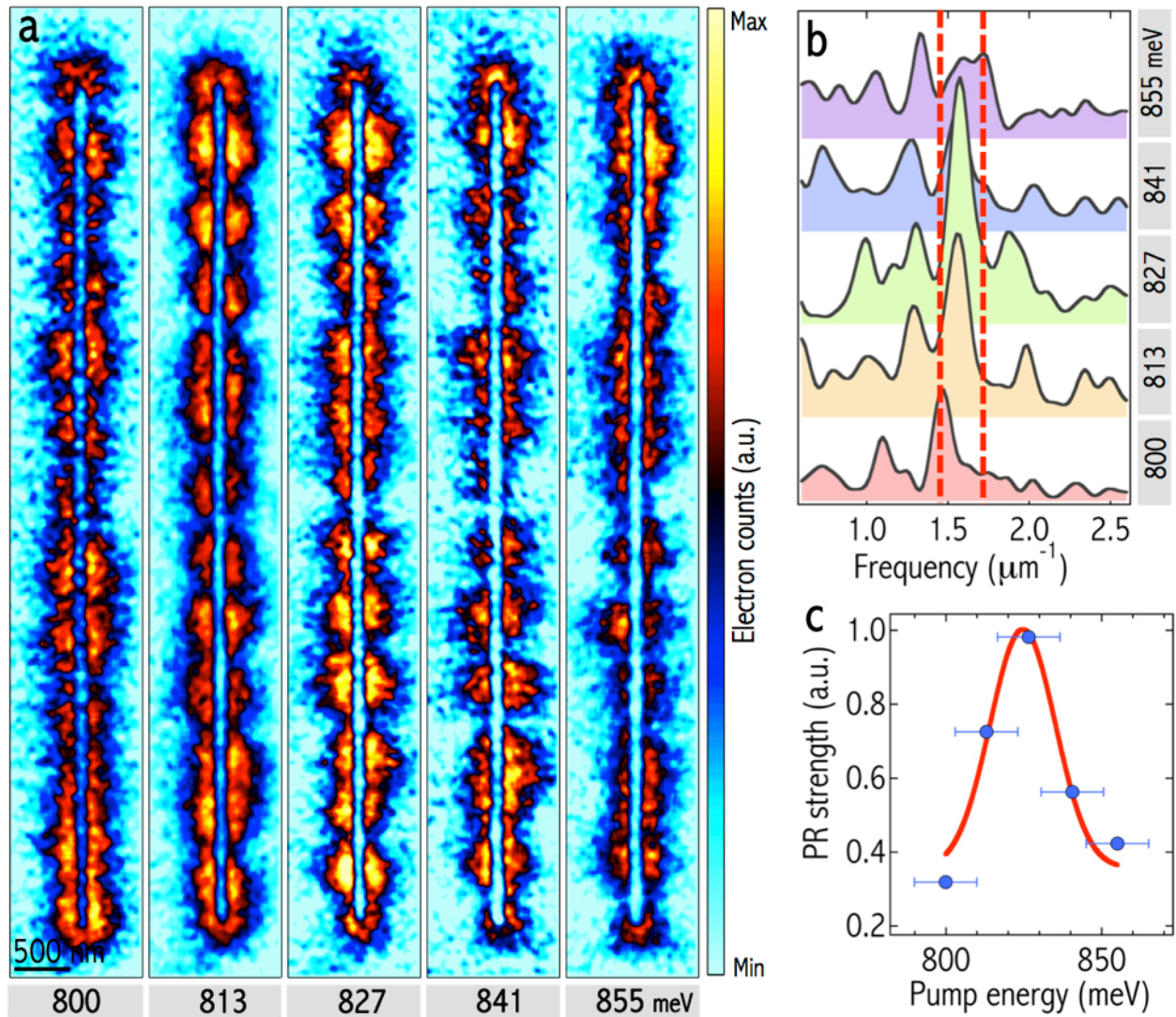

Fig. 3 ***Laser-assisted plasmon spectroscopy***. ***a*** TEM images of the n = 13 plasmon resonance for selected photo-excitation energies around 827 meV (1500 nm, central image). ***b*** Fourier transform (FT) of the parallel-to-the-wire spatial profile in these images as a function of laser energy. ***c*** Integral of the pump-energy-dependent FT within the region of spatial frequencies corresponding to the mode n = 13 (indicated by red-dashed lines in ***b***).

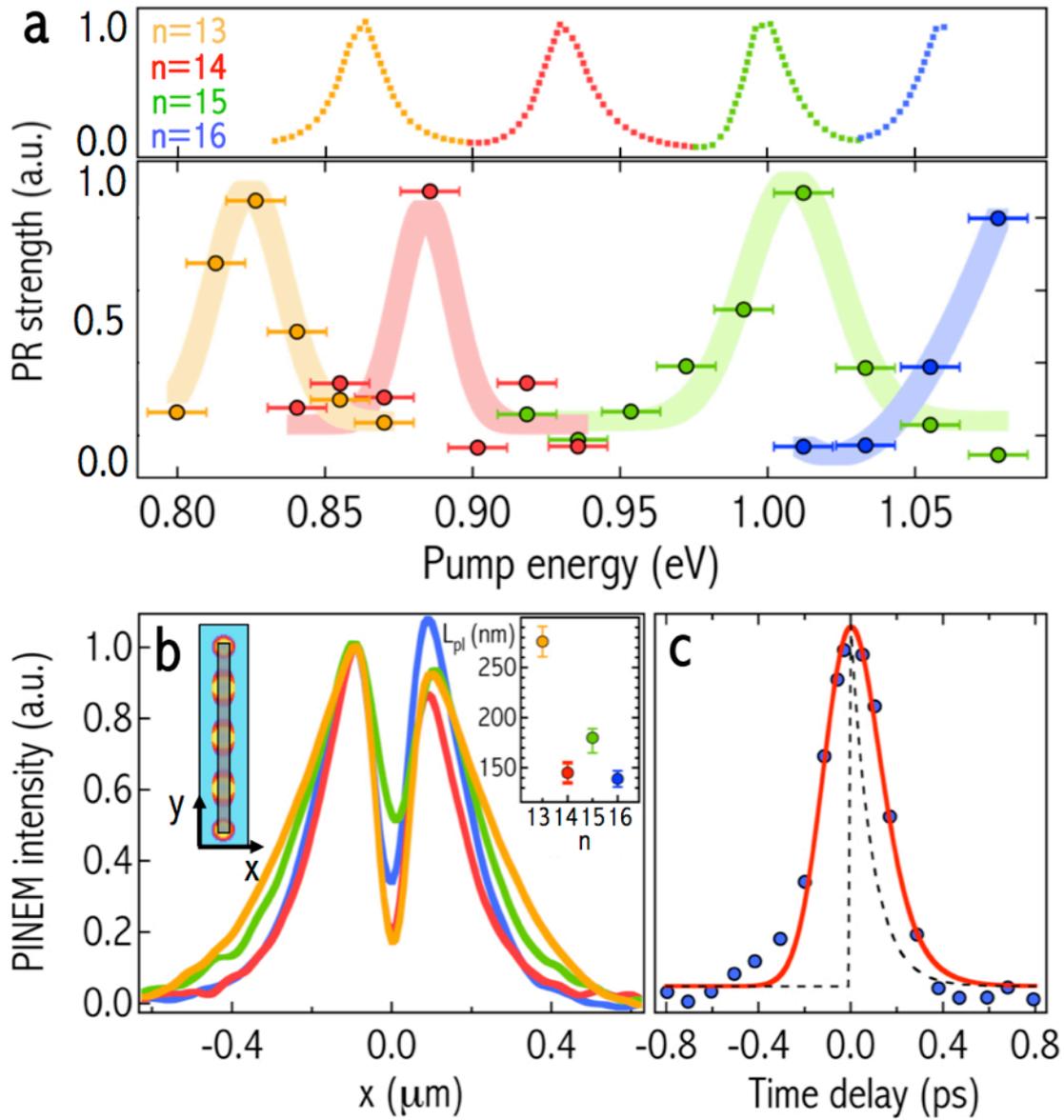

Fig. 4 *Energy, space, and time mapping of plasmons.* **a** *Plasmon spectra for modes n = 13-16 (see color-coded labels) obtained upon scanning the laser energy between 0.8 (1550 nm) and 1.08 eV (1150 nm). Symbols are obtained with the integrated FT procedure explained in Fig. 3. Dashed curves on the top correspond to simulations of the PRs strengths for a nanowire of 8 μm length and 100 nm diameter.* **b** *Spatial profile of the photo-excited nanowire plasmonic field along the transversal x axis (left inset), obtained after integrating along the y wire axis for the different modes under consideration. Right inset: spatial variation of the plasmonic fields as a function of*

*mode index n.* ***c*** *Temporal evolution of the photo-excited plasmonic field (blue circles), fitted with an exponential decay (dashed-black curve) convoluted with a Gaussian IRF (red curve).*

**SUPPLEMENTARY FIGURES**

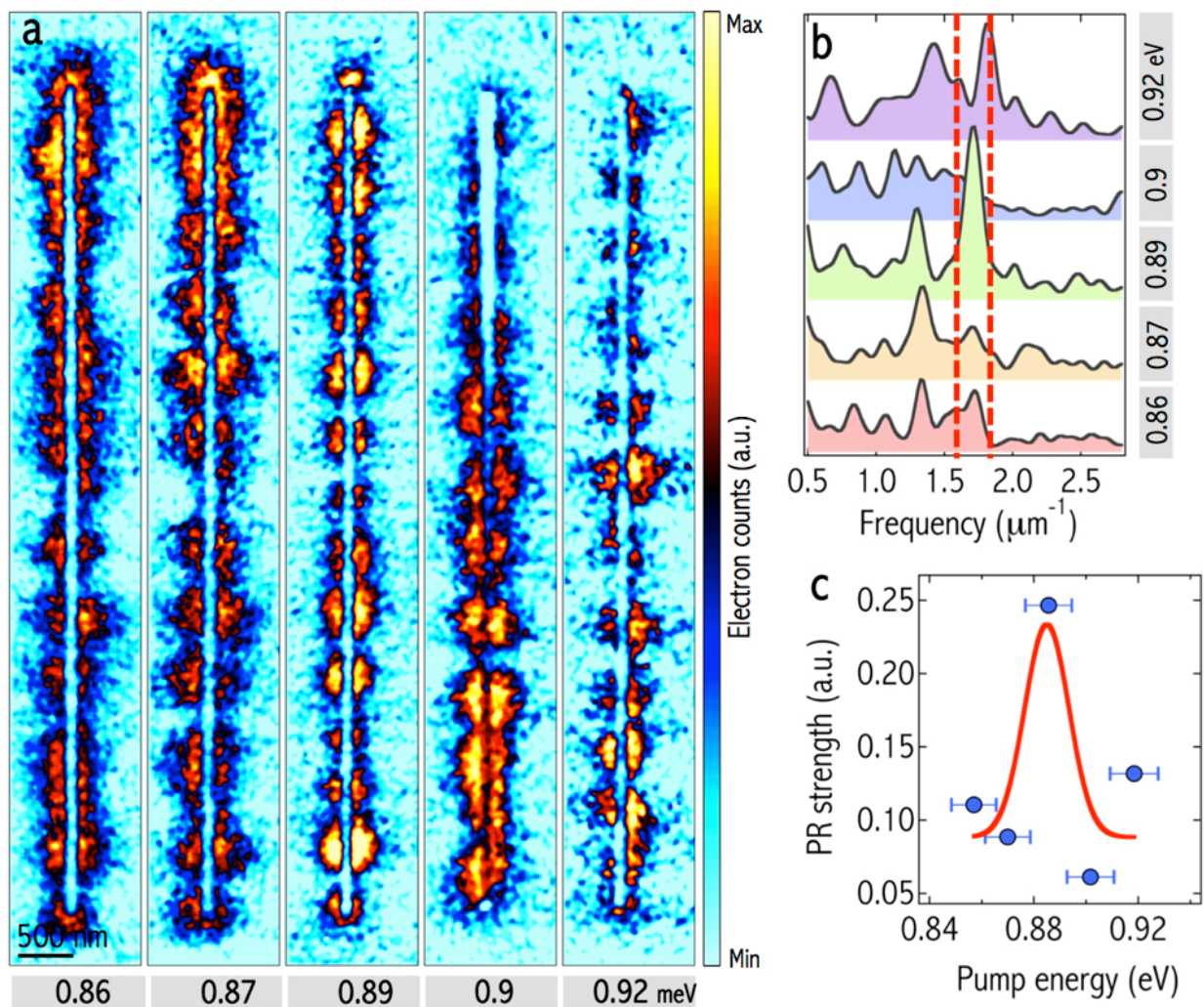

Fig. S1 *Same as Fig. 3 for the n = 14 plasmon resonance.*

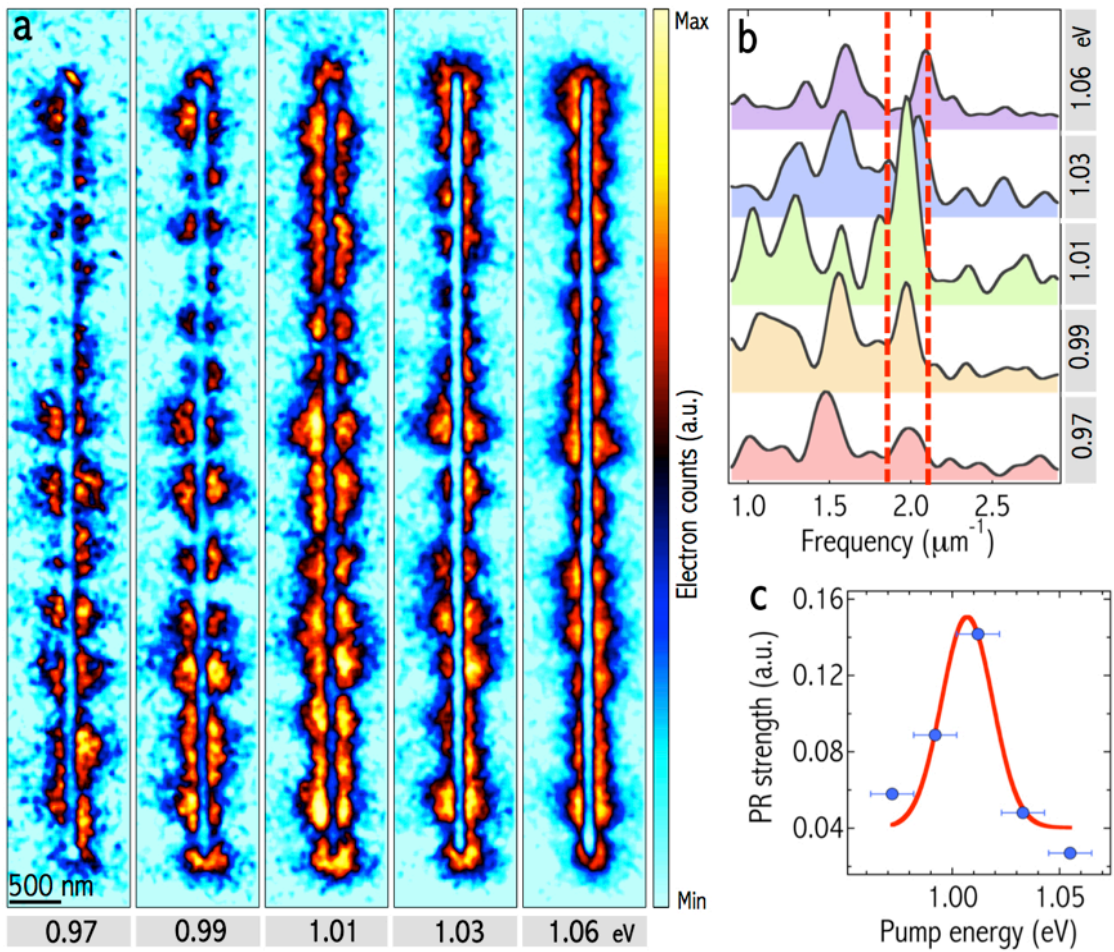

Fig. S2  *Same as Fig. 3 for the n = 15 plasmon resonance.*

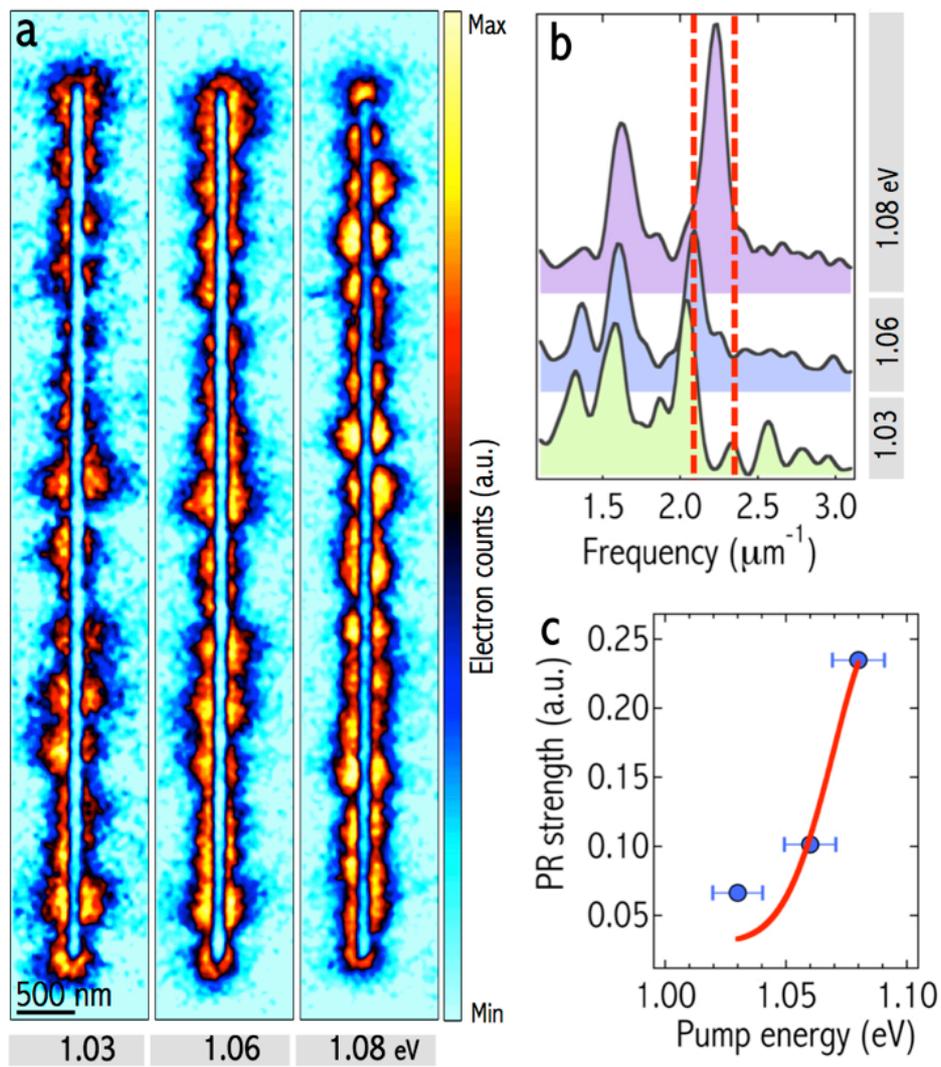

Fig. S3  *Same as Fig. 3 for the n = 16 plasmon resonance.*